# Developing signal processing skills: A proposal for the teaching of filters.

Eng. Fernanda Zapata Bascuñán[1], Msc. Eng. Marcelo Aráoz[2] and Eng. Daniel Colón[2]

[1] Escuela de Ciencia y Tecnología, Universidad Nacional de San Martín, Buenos Aires.
[2] Facultad de Ingeniería, Universidad Nacional del Comahue, Neuquén.
zapatafer945@gmail.com

**Abstract.** In accordance with Bloom's taxonomy, a four-level evaluation abstraction was generated with the objective of structuring and hierarchizing curricula knowledge, allowing students to dominate a subject and progressively reach the top of Bloom's pyramid. The evaluation was conducted with 10 university students who analyzed and optimized a filtering process using MATLAB simulation while constrained by time. The results demonstrate that this novel teaching method facilitates a more profound comprehension of the subject for students with diverse learning styles in science.

# Desarrollo de competencias en procesamiento de señales: una propuesta didáctica para la enseñanza de filtros.

**Resumen.** En concordancia con la taxonomía de Bloom, se desarrolló un marco de evaluación de cuatro niveles con el objetivo de estructurar y jerarquizar el conocimiento curricular, permitiendo que los estudiantes dominen una asignatura y progresivamente alcancen los niveles cognitivos superiores de la pirámide de Bloom. La evaluación se llevó a cabo con diez estudiantes universitarios que analizaron y optimizaron un proceso de filtrado mediante simulaciones en MATLAB, bajo restricciones de tiempo. Los resultados indican que este enfoque didáctico innovador favorece una comprensión más profunda del tema, adaptándose a diversos estilos de aprendizaje entre los estudiantes de ciencias.

**Palabras clave:** Taxonomía de Bloom, Educación en Procesamiento de Señales, Pedagogía Inclusiva, Desarrollo de Habilidades Cognitivas.

## 1 Introduction

Today, the electronics industry demands a deep understanding not only of technological tools, but also of their operational context and intrinsic constraints. In this sense, competence-based teaching emerges as a vital pedagogical approach to prepare higher education students for the challenges of the world of work.

This paper explores this intersection between technological competences in a higher education context by exploring a pedagogical strategy that uses simulation as a tool. Simulation not only allows students to experiment with theoretical concepts in a controlled environment, but also gives them the opportunity to address the practicalities and limitations of the real world. To structure the analysis of these competences in the context of simulation, we use Bloom's taxonomy, which provides a comprehensive framework for assessing and developing different levels of cognitive skills.

Through this approach, this work aims not only to impart technical knowledge, but also to cultivate transferable skills and a greater understanding of digital signal processing.

### 1.1 Competence framework

It is essential to explore the conceptual framework underlying the concept of competences. It is recognized that there is a wide range of interpretations around this term, but generally speaking, there is an emerging consensus on the key elements that comprise it. Competence can be conceptualized as the holistic manifestation of an individual, encompassing not only concrete and descriptive knowledge, but also capacities, skills, attitudes and principles, framed in an ethical and practical context.

This complex notion of competence highlights the importance of understanding three fundamental aspects of learning: '*knowing how to be*', '*knowing how to do*' and '*knowing how to know*'. These three interdependent pillars provide a complete and holistic view of what the learner is expected to achieve. As Pimienta Prieto points out in his book [1], this approach considers the teacher not only as a transmitter of knowledge, but as a professional who acts as a mediator and catalyst of the learning process. On the other hand, the student is conceived as an active agent in the construction of his or her own integral education, adopting a leading role in the acquisition and application of competences.

This perspective emphasize the interaction between teacher and student, where the former guides and encourages the development of skills and abilities, while the latter takes responsibility for his or her own learning process. Ultimately, this view of competence-based learning places both teachers and students in a collaborative and enriching role.

## 2 Description of the experience

### 2.1 Context in which the activity is set

The activity is implemented as a follow-up to the didactic sequence on the Fourier transform. The activity is presented as a structured group guide, aimed at breaking down in detail the process of applying the Fourier transform in four different contexts: discrete or continuous signals, with periodic or aperiodic properties.

The guide is characterized by its arrangement in gradual levels of complexity, which allows students to progressively advance in their understanding and application of the Fourier transform. In addition, it helps in the construction of a simulation environment, providing a structure that can serve as a model for other activities that the student faces.

Another fundamental aspect to highlight is that this guide is solidly based on the central reference book for the subject, '*Applied Digital Signal Processing*' [5].

The following activity follows the Fourier transform activity and is linked to signal processing in the spectrum: the implementation of a filter. In the following sections, the activity and the expectations surrounding it will be detailed.

### 2.2 Activity description

As usual in various courses related to discrete and continuous time signal processing, we specifically address the implementation of filters in a controlled environment. For this task, we take advantage of the MATLAB computer system. This platform has established itself as a valuable resource for this activity and has led to the generation of numerous publications of interest, including the work of Barrus [2, 3, 4].

The task given to the student is quite simple: they are told that a student has filtered the signal in order to eliminate the noise it contains but has not achieved it in an optimal way. They must submit a report in which they describe the analysis carried out by the previous designer (1), the analysis and criteria that led them to determine the error that the previous designer made (2) and state the criteria adopted to arrive at the optimal solution (3). If they did not reach the optimal solution, they should state what it would have been and the reasons that prevented them from reaching it.

In addition, they were limited to one and a half hours to solve the problem.

The students, at the moment of developing this activity, have a solid knowledge in spectral analysis by means of the Fourier transform, which allows them to analyze the signal and determine the type of noise associated to the signal or to make a contrast between the original signal (which contained noise) and the signal filtered by the previous user, so it is possible to characterize the filter. In addition to other potential tools that are provided by the computational system, such as *sptool* or *filter design* [5].

In this context, we consider that this methodology offers an alternative to assess a set of competences related to the subject. Students demonstrate knowledge by recog-nizing the environment, evaluating the mistakes made by others and identifying the optimal tool, knowing how to know (1).

Through the implementation of the filter, knowledge of the computational tool and spectral analysis, know-how (2).

Attitudinally, they are facing a challenge against the clock, know how to be (3).

### 2.3 Levels of abstraction involved and assessment

Simulation is a process that can be categorized into different steps, activities or roles to be developed in order to perform optimally, as is the case in [6]. If we consider that this activity can be analyzed as levels of abstraction, we can relate it almost directly to Bloom's taxonomy [7].

For practical purposes we will analyze the proposed simulation activity under this perspective, relating each step necessary for the development of the activity to a level of abstraction.

First, we start by recognizing the information, which involves retrieving relevant knowledge from long-term memory and comparing it with the information presented. This activity falls into the lowest level of abstraction proposed by Bloom, as it involves recalling previously acquired information and applying it.

The inference step rises to the level of '*Comprehension*', as it involves not only remembering, but also understanding and relating information in a specific context. Inference represents the ability to compare and contrast information to deduce an error or deficiency in the simulation process.

The attribution stage addresses the '*Analysis*' level of the taxonomy, as it requires breaking down and examining the material presented to understand the underlying intentions. Here, the student goes beyond comprehension, drawing out implications and possible motivations of the author of the material.

The attribution report, which culminates in the presentation of the activity report, aligns with the '*Synthesis*' level in Bloom's taxonomy. It requires the creation of a final product that combines the attributions made and sets them out coherently, demonstrating the learner's ability to integrate information and generate new knowledge from it.

Finally, the explicitness of the criteria used to solve the detected deficiency corresponds to the '*Evaluation*' level. The student engages in an informed decision-making process, justifying their actions and demonstrating a thorough understanding of the relevant concepts and methods.

Ultimately, the proposed assessment activity not only promotes practical understanding of signal processing concepts, but also cultivates critical and analytical skills essential for students as they move through their educational training.

## 3 Student Performance and relation with the criteria

### 3.1 Evaluation rubric and observed

In order to assess the effectiveness of the improved methodology, a comparison was made between the outcomes and those from previous evaluations involving the same participants.

The following table presents the results of the evaluation.

Table 1. Results of evaluation with and without Bloom's taxonomy criteria.

| Students | C1 (without Bloom's taxonomy) | C2 (without Bloom's taxonomy) | REC C (with Bloom's taxonomy) | Best Grade (C1 or C2) | Estimated improvement (Points) | Percentage of Improvement |
|---|---|---|---|---|---|---|
| 1st | 5 | 6 | 15 | 6 | 4 | 20% |
| 2nd | 2 | 3 | 12 | 3 | 7 | 35% |
| 3th | 1.5 | 2 | 16 | 2 | 12.5 | 62.5% |
| 4th | 5 | 3 | 12 | 5 | 4 | 20% |
| 5th | Aus | 5 | 14 | 5 | 9 | 45% |
| 6th | Just | 2 | 12 | 2 | 10 | 50% |
| 7th | 1 | Aus | 12 | 1 | 11 | 55% |
| 8th | 2.5 | 6 | 15 | 6 | 6.5 | 32.5% |
| 9th | 6 | 3 | 15 | 6 | 6 | 30% |
| 10th | 3,5 | 8.5 | 16.5 | 8.5 | 4.5 | 22.5% |

The detailed evaluation criteria are provided in the annex, where the point distribution for each component is clearly outlined. Although individual scores for each aspect were not preserved—due to the fact that they were not digitized or properly stored—the overall assessment results show an improvement of up to 62.5% compared to the previous evaluation. In some cases, the improvement was no more than 20%, with an average increase of 37.5%.

The improvement was analyzed as follows: evaluations C1 and C2 each had a maximum score of 10 points, while the REC C2 assessment had a maximum of 20 points, as it integrated the curricular content from both C1 and C2. To calculate the improvement, the points obtained in REC C2 were compared against the combined scores of C1 and C2. The improvement in points was calculated as:

Improvement = (REC C2 score – (C1 + C2))

This difference was then divided by the maximum score of 20 and multiplied by 100 to determine the percentage improvement in evaluation metrics:

% Improvement = [(REC C2 − (C1 + C2)) / 20] × 100

This approach allowed for a standardized comparison of student progress across the assessments.

## 4 Observations in the Activity Design and Implementation

While student engagement and performance were strong across cognitive domains, certain limitations in the activity design were identified:

The absence of a meaningful title may have hindered initial comprehension of the activity's objectives, potentially impacting student performance at the Knowledge and Comprehension levels of Bloom's taxonomy. Although students were given an opportunity to ask questions about the task instructions, it is strongly believed that providing a more descriptive and informative title for each activity would enhance understanding from the outset.

Additionally, the limited time-frame likely constrained the depth of analysis and the quality of synthesis and evaluation, particularly for students operating at higher levels of abstraction. This factor should be carefully considered when designing activities from scratch. In particular, the nearly two-hour time allocation may be insufficient for students with learning differences—such as dyslexia—who may require additional time to process information and complete complex cognitive tasks [9], or a time frame better aligned with their processing needs.

## 5 Recommendations for Future Implementation

To foster a more enriching learning experience across all cognitive levels, future implementations should incorporate:
- *Hardware devices* to reinforce practical understanding and extend the application of concepts.
- *Design constraints* (e.g., rounding, cutoff) to encourage deeper analysis and evaluation in digital filter design.
- *A more descriptive title and clearer objectives* to enhance comprehension and support progression through Bloom's levels more effectively.
- A clear description of the time required for each task. This approach relieves students from having to manage time themselves and supports those with time blindness or dyscalculia in better navigating their assessment.

## 6    Conclusion

During the course of the activity, students demonstrated a clear aptitude for accurately discerning the intentions of the preceding designer. Notably, they were able to develop effective and contextually appropriate solutions to the challenges presented. A particular strength observed was their autonomy in formulating these solutions, which were well aligned with both the environment and the nature of the task. Furthermore, students' ability to articulate the reasoning and processes behind their decisions was evaluated, highlighting their depth of understanding and reflective capacity.

It is important to note that the structured evaluation strategy—grounded in levels of abstraction-provides students with a clear framework for understanding and navigating the learning process. Maintaining consistent assessment criteria throughout the course is essential; therefore, it is advisable to avoid modifying the evaluation scheme mid-course.

Moreover, this form of assessment implicitly supports not only the enhancement of student performance but also enables instructors to more effectively gauge varying levels of understanding among students. In particular, it offers valuable insights for *identifying and supporting neurodivergent students or those requiring additional academic assistance*. By adopting this approach, we can foster inclusive pedagogical practices and contribute to the development of diverse cognitive skill sets.

In light of these findings, it is recommended that future implementations of this activity be carried out with an even more refined and enriching perspective. Considerations may include the integration of additional technological tools or refined design constraints to further support student learning and critical engagement.

| Dimension | Excellent (4) | Proficient (3) | Basic (2) | Needs Improvement (1) |
|---|---|---|---|---|
| **Recognition and Application** (Knowledge & Comprehension) | Accurately retrieves and integrates relevant prior knowledge; effectively compares and applies it to new information. | Retrieves and applies prior knowledge appropriately; comparisons are generally clear. | Demonstrates limited ability to recall or apply relevant knowledge; comparisons are vague or underdeveloped. | Fails to retrieve or apply relevant prior knowledge meaningfully. |
| **Inference** (Comprehension) | Clearly identifies contextual relationships and accurately infers intentions and errors within the simulation, supported by sound reasoning. | Makes appropriate inferences and identifies relationships with minor gaps in clarity or detail. | Inferences are incomplete or weakly supported; contextual understanding is limited. | Does not identify key relationships or infer intentions effectively. |
| **Attribution** (Analysis) | Independently analyzes complex aspects and thoughtfully examines the motivations behind the design decisions. | Provides a generally accurate analysis and identifies basic motivations with some support. | Offers a limited or surface-level analysis; motivations are mentioned but not explored. | Lacks meaningful analysis or insight into design motivations. |
| **Synthesis and Communication** (Synthesis) | Integrates knowledge cohesively; final report is original, clearly structured, and effectively communicates new insights. | Synthesizes information logically and communicates clearly, though with minor issues in depth or cohesion. | Demonstrates partial synthesis; communication lacks clarity or organization. | Fails to synthesize information or communicate ideas coherently. |
| **Evaluation and Justification** (Evaluation) | Provides well-reasoned justifications and proposes criteria-based, insightful solutions. | Offers reasonable justifications and proposes generally suitable solutions. | Justifications are weak or unclear; solutions lack strong rationale. | Does not provide adequate justification or propose coherent solutions. |